\documentclass[11pt]{article}

\usepackage{epsfig,latexsym}
\usepackage{amsmath}

\newcommand{\join}{\text{\textcircled{{\footnotesize 1}}}}
\newcommand{\cojoin}{\text{\textcircled{{\footnotesize 0}}}}

\newcommand{\qed}{\hfill $\Box$}

\newtheorem{defi}{Definition}
\newtheorem{fact}{Fact}
\newtheorem{theo}{Theorem}
\newtheorem{lemm}{Lemma}
\newtheorem{coro}{Corollary}

\newtheorem{clai}{Claim}[section]

\begin{document}

\author{
Andreas Brandst\"adt\\
\small Institut f\"ur Informatik, Universit\"at Rostock, D-18051 Rostock, Germany\\
\small \texttt{andreas.brandstaedt@uni-rostock.de}\\
\and
Raffaele Mosca\\
\small Dipartimento di Economia, Universit\'a degli Studi ``G. D'Annunzio'', 
Pescara 65121, Italy\\
\small \texttt{r.mosca@unich.it}
}

\title{Maximum Weight Independent Sets for $(S_{1,2,4}$,Triangle$)$-Free Graphs in Polynomial Time}

\maketitle

\begin{abstract}
The Maximum Weight Independent Set (MWIS) problem on finite undirected graphs with vertex weights asks for a set of pairwise nonadjacent vertices of maximum weight sum. MWIS is one of the most investigated and most important algorithmic graph problems; it is well known to be NP-complete, and it remains NP-complete even under various strong restrictions such as for triangle-free graphs. Its complexity for $P_k$-free graphs, $k \ge 7$, is an open problem. In \cite{BraMos2018}, it is shown
that MWIS can be solved in polynomial time for ($P_7$,triangle)-free graphs. This result is extended by Maffray and Pastor 
\cite{MafPas2016} showing that MWIS can be solved in polynomial time for ($P_7$,bull)-free graphs. In the same paper, they also showed that MWIS can be solved in polynomial time for ($S_{1,2,3}$,bull)-free graphs.

In this paper, using a similar approach as in \cite{BraMos2018}, we show that MWIS can be solved in polynomial time for ($S_{1,2,4}$,triangle)-free graphs which generalizes the result for ($P_7$,triangle)-free graphs. 

\end{abstract}


\section{Introduction}

Let $G$ be a finite, simple and undirected graph and let $V(G)$ (respectively, $E(G)$) denote the vertex set (respectively, the edge set) of $G$. 
For $U \subseteq V(G)$, let $G[U]$ denote the subgraph of $G$ induced by $U$. Throughout this paper, all subgraphs are understood as induced subgraphs. 

For $v \in V(G)$, let $N(v) := \{u \in V(G) \setminus \{v\} : uv \in E(G)\}$ be the {\em open neighborhood} of $v$ in $G$, let $N[v]:=N(v) \cup \{v\}$ be the {\em closed neighborhood} of $v$ in $G$, and let $A(v):=V(G) \setminus N[v]$ be the {\em anti-neighborhood} of $v$ in $G$. For $v \in V(G)$ and $U \subseteq V(G)$, with $v \not \in U$, let $N_U(v) := N(v) \cap U$. 

If $u \in N(v)$ ($u \notin N(v)$, respectively) we say that {\em $u$ sees $v$} ({\em $u$ misses $v$}, respectively).  
An {\em independent set} (or {\em stable set}) in a graph $G$ is a subset of pairwise nonadjacent vertices of $G$. An independent set in a graph $G$ is {\em maximal} if it is not properly contained in any other independent set of $G$.

Given a graph $G$ and a weight function $w$ on $V(G)$, the {\em Maximum Weight Independent Set} ({\em MWIS}) {\em problem} asks for an independent set of $G$ with maximum weight. Let $\alpha_w(G)$ denote the maximum weight of an independent set of $G$. The MWIS problem is called {\em MIS problem} if all vertices $v$ have the same weight $w(v) = 1$.

The MIS problem ([GT20] in \cite{GarJoh1979}) is well known to be NP-complete \cite{Karp1972}. While it is solvable in polynomial time for bipartite graphs (see 
e.g.\ \cite{AhuMagOrl1993,DesHak1970,GroLovSch1988}), it remains NP-hard even under various strong restrictions, such as for triangle-free graphs \cite{Polja1974}. 

The following specific graphs are subsequently used. $P_k$ has vertices $v_1,v_2,\ldots,v_k$ and edges $v_jv_{j+1}$ for $1 \le j < k$. $C_k$ has vertices $v_1,v_2,\ldots,v_k$ and edges $v_jv_{j+1}$ for $1 \le j \le k$ (index arithmetic modulo $k$). $K_{\ell}$ has $\ell$ vertices which are pairwise adjacent. Clearly, $C_3=K_3$. $C_3$ (and thus, $K_3$) is also called {\em triangle}.
A {\em claw} (with center $a$) has vertices $a,b,c,d$ and edges $ab,ac,ad$. $S_{i,j,k}$ (with center $a$) is the graph obtained from a claw with center $a$ by subdividing respectively its edges into $i$, $j$, $k$ edges (e.g., $S_{0,1,2}$ is a $P_4$, $S_{1,1,1}$ is a claw).

For a given graph $F$, a graph $G$ is {\em $F$-free} if no induced subgraph of $G$ is isomorphic to $F$. If for given graphs $F_1,\ldots,F_k$, $G$ is $F_i$-free for all $1 \le i \le k$ then we say that $G$ is {\em $(F_1,\ldots,F_k)$-free}.

Alekseev \cite{Aleks1983,Aleks2004/2} proved that, given a graph class ${\cal{X}}$ defined by forbidding a finite family $\cal{F}$ of induced graphs, the MIS problem remains NP-hard for the graph class ${\cal{X}}$  if each graph in $\cal{F}$ is not an $S_{i,j,k}$ for some index $i,j,k$. Various authors \cite{FaeOriSta2011,Minty1980,NakTam2001,NobSas2015,Sbihi1980} proved that MWIS can be solved for claw-free (i.e., $S_{1,1,1}$-free) graphs in polynomial time (improving the time bounds step by step). 
Lozin and Milani\v c \cite{LozMil2008} proved that MWIS can be solved for fork-free graphs (i.e., $S_{1,1,2}$-free graphs) in polynomial time $-$ Alekseev \cite{Aleks1999,Aleks2004/1} previously proved a corresponding result for the unweighted case. 

In this paper, we show that for ($S_{1,2,4}$,triangle)-free graphs, MWIS can be solved in polynomial time. This generalizes the polynomial-time result for MWIS on 
($P_7$,triangle)-free graphs \cite{BraMos2018} (which was extended by Maffray and Pastor \cite{MafPas2016} showing that MWIS can be solved in polynomial time for ($P_7$,bull)-free graphs; in the same paper, they also showed that MWIS can be solved in polynomial time for ($S_{1,2,3}$,bull)-free graphs).   

The following result is well known:

\begin{theo}[\cite{AhuMagOrl1993,DesHak1970,GroLovSch1988}]\label{MWSbipgrpoltime}
Let $B$ be a bipartite graph with $n$ vertices.
\begin{enumerate}
\item[$(i)$]  MWIS $($with rational weights$)$ is solvable for $B$ in time ${\cal O}(n^4)$ via linear programming
              or network flow.
\item[$(ii)$] MIS is solvable for $B$ in time ${\cal O}(n^{2.5})$.   
\end{enumerate}
\end{theo}

A graph $G$ is {\em nearly bipartite} if, for each $v \in V(G)$, the subgraph $G[A(v)]$ induced by its anti-neighborhood is bipartite. Obviously we have:
\begin{equation}\label{antineighbapproach}
\alpha_w(G) = \max_{v \in V(G)} \{w(v)+\alpha_w(G[A(v)])\}
\end{equation}
Thus, by Theorem \ref{MWSbipgrpoltime}, the MWIS problem (with rational weights) can be solved in time ${\cal O}(n^5)$ for nearly bipartite graphs.

Our approach is based on a repeated application of the {\em anti-neighborhood approach} with respect to (\ref{antineighbapproach}) (and in particular, on the approach for MWIS on ($P_7$,triangle)-free graphs \cite{BraMos2018}).   

That allows, by detecting an opportune sequence of vertices, to split and to finally reduce the problem to certain instances of bipartite subgraphs, for which the problem can be solved in polynomial time (recall Theorem \ref{MWSbipgrpoltime}). In particular, as a corollary we obtain: For every 
($S_{1,2,4}$,triangle)-free graph $G$ there is a family ${\cal S}$ of subsets of $V(G)$ inducing bipartite subgraphs of $G$, with ${\cal S}$ detectable in polynomial time and containing polynomially many members, such that every maximal independent set of $G$ is contained in some member of ${\cal S}$. That seems to be harmonic to the result of Pr\"omel et al.\ \cite{ProSchSte2002} showing that with ``high probability'', removing a single vertex in a triangle-free graph leads to a bipartite graph.

\subsection{Further notations and preliminary results}

For any missing notation or reference let us refer to \cite{BraLeSpi1999}.
For $U,W \subseteq V(G)$, with $U \cap W = \emptyset$, $U$ has a {\em join} (a {\em co-join}, respectively) to $W$, denoted by $U \join W$ ($U \cojoin W$, respectively), if each vertex in $W$ is adjacent (is nonadjacent, respectively) to each vertex in $U$.

For $v \in V(G)$ and $U \subseteq V(G)$, with $v \not \in U$, $v$ {\em contacts} $U$ if $v$ is adjacent to some vertex of $U$; 
$v$ {\em dominates} $U$ if $v$ is adjacent to all vertices of $U$, that is, $\{v\} \join U$ ($v \join U$ for short); 
$v$ {\em misses} $U$ if $v$ is non-adjacent to all vertices of $U$, that is, $\{v\} \cojoin U$ ($v \cojoin U$ for short).

A {\em component of $G$} is a maximal connected subgraph of $G$. 
The {\em distance} $d_G(u,v)$ of two vertices $u,v$ in $G$ is the number of edges of $G$ in a shortest path between $u$ and $v$ in $G$.

%For a subgraph $H$ of $G$ and $k \ge 0$, a vertex $v \notin V(H)$ is a {\em $k$-vertex for $H$} (or {\em of $H$}) if it has exactly
%$k$ neighbors in $H$. {\em $H$ has no $k$-vertex} if there is no $k$-vertex for $H$.

Recall that $C_3$ is a triangle.

\begin{lemm}\label{S124C3C5frnearlybip}
Connected $(S_{1,2,4},C_3,C_5)$-free graphs are nearly bipartite.
\end{lemm}

{\bf Proof.} Let $G=(V,E)$ be a connected $(S_{1,2,4},C_3,C_5)$-free graph. Suppose to the contrary that for some vertex $v$, $G[A(v)]$ is not bipartite, i.e., $G[A(v)]$ contains an odd chordless cycle. Then, since $G$ is ($C_3,C_5$)-free, $G[A(v)]$ contains an odd chordless cycle $C_{2k+1}$, say $C$, for some $k \geq 3$. Let $\{v_1,\ldots,v_{2k+1}\}$ be the vertices of $C$ and let $v_iv_{i+1}$ (index arithmetic modulo $2k+1$) be the edges of $C$. Then let $P$ be a shortest path between $v$ and $C$; clearly, the distance between $v$ and $C$ is at least 2. Without loss of generality (since the other cases can be similarly treated), assume that $P$ has exactly one internal vertex, say $d$ (i.e., $d$ is adjacent to $v$ and to some vertex of $C$). 

\begin{clai}\label{iimpliesi+2ori+4}
If $dv_i \in E$ then $dv_{i+2} \in E$ or $dv_{i+4} \in E$.  
\end{clai}

{\em Proof.}
Since $G$ is $C_3$-free, $dv_i \in E$ implies $dv_{i-1} \notin E$ and $dv_{i+1} \notin E$, and since $G$ is $C_5$-free and thus, $d,v_i,v_{i+1},v_{i+2},v_{i+3}$ do not induce a $C_5$, we have $dv_{i+3} \notin E$. Now, since $v_i,v_{i-1},d,v,v_{i+1},v_{i+2},v_{i+3},v_{i+4}$ (with center $v_i$) do not induce an $S_{1,2,4}$, 
we have $dv_{i+2} \in E$ or $dv_{i+4} \in E$ which shows Claim \ref{S124C3C5frnearlybip}.
$\diamond$

Now, since $C$ is an odd cycle, Claim \ref{iimpliesi+2ori+4} leads to a $C_5$ or $C_3$ which is a contradiction (for example, if $C=C_7$ and $dv_1 \in E$ then clearly, $dv_7 \notin E$, and $dv_5 \in E$ leads to a $C_5$ with vertices $d,v_5,v_6,v_7,v_1$). 

Thus Lemma \ref{S124C3C5frnearlybip} is shown. 
\qed

\medskip

Since by Lemma \ref{S124C3C5frnearlybip}, every component of a $(S_{1,2,4},C_3,C_5)$-free graph $G$ is nearly bipartite, 
and since MWIS is solvable in polynomial time for nearly bipartite graphs (recall Theorem \ref{MWSbipgrpoltime} and MWIS for nearly bipartite graphs), we have: 

\begin{coro}\label{MWISS124K3C5fr}
MWIS is solvable in polynomial time for $(S_{1,2,4},C_3,C_5)$-free graphs.
\end{coro}

Our aim is to show that MWIS can be solved in polynomial time for ($S_{1,2,4},C_3$)-free graphs. Since by Corollary \ref{MWISS124K3C5fr}, we are done with $(S_{1,2,4},C_3,C_5)$-free graphs, from now on let $G$ be a connected ($S_{1,2,4},C_3$)-free graph containing a $C_5$.
Using again the anti-neighborhood approach, let $v \in V(G)$ and let $K$ be a component of the induced subgraph $G[A(v)]$ of its anti-neighborhood $A(v)$.
Since $G$ is connected, $v$ has a neighbor $w \in N(v)$ contacting $K$. Since $G$ is $C_3$-free, $N(v)$ is independent.

A component $T$ of $G[Z]$ is {\em nontrivial} if $T$ contains a $P_2$. 

\begin{fact}\label{S124C3frtwoK}
For any $(S_{1,2,4},C_3)$-free graph $G=(V,E)$ and for any $v \in V$ and its anti-neighborhood $A(v)$, if $w \in N(v)$ contacts two nontrivial components $K,K'$ of $G[A(v)]$ then for any $P_4$ $(x_1,x_2,x_3,x_4)$ in $K$, if $wx_1 \in E$ then $wx_3 \in E$ or $wx_4 \in E$.
\end{fact}

{\bf Proof.} 
Let $(y_1,y_2)$ be a $P_2$ in $K'$ which is contacted by $w$, say $wy_1 \in E$, and clearly, $wy_2 \notin E$ since $G$ is $C_3$-free. For a $P_4$ $(x_1,x_2,x_3,x_4)$ in $K$, let $wx_1 \in E$. Then, since $w,v,y_1,y_2,x_1,x_2,x_3,x_4$ (with center $w$) do not induce an $S_{1,2,4}$, we have $wx_3 \in E$ or $wx_4 \in E$. 
Thus, Fact \ref{S124C3frtwoK} is shown. 
\qed

\begin{lemm}\label{S124C3fratmostoneKwithC5}
For any $(S_{1,2,4},C_3)$-free graph $G=(V,E)$ and for any $v \in V$ and its anti-neighborhood $A(v)$, at most one component $K$ of $G[A(v)]$ can contain a $C_5$.
\end{lemm}

{\bf Proof.} Let $G=(V,E)$ be a connected $(S_{1,2,4},C_3)$-free graph. Suppose to the contrary that for some vertex $v$, there are two components $K,K'$ in $G[A(v)]$ containing a $C_5$, say $C=(x_1,x_2,x_3,x_4,x_5)$ in $K$ and $C'=(y_1,y_2,y_3,y_4,y_5)$ in $K'$. Let $w \in N(v)$ be a neighbor of $v$ contacting $K$, and let $w' \in N(v)$ be a neighbor of $v$ contacting $K'$. First assume that $w$ contacts $K$ and $K'$. Clearly, $K$ and $K'$ are nontrivial. By Fact \ref{S124C3frtwoK}, $w$ contacts $C$ since otherwise, there is a $P_4$ $P$ in $K$ such that $w$ contacts only one end-vertex of $P$, and correspondingly, $w$ contacts $C'$ in $K'$. 
Without loss of generality, let $wx_1 \in E$ and $wy_1 \in E$. Then again by Fact \ref{S124C3frtwoK}, $w$ has exactly two neighbors in $C$ and in $C'$, say $wx_3 \in E$ and $wy_3 \in E$. But now, $x_3,x_2,x_4,x_5,w,y_1,y_5,y_4$ (with center $x_3$) induce an $S_{1,2,4}$ in $G$ which is a contradiction. 

Thus, no neighbor $w \in N(v)$ contacts $K$ and $K'$; let $w \in N(v)$ contact $K$ and let $w' \in N(v)$ contact $K'$, $w \neq w'$, while $w \cojoin V(K')$ and $w' \cojoin V(K)$. Recall that $N(v)$ is independent, i.e., $ww' \notin E$. Without loss of generality, let $wx_1 \in E$ and $w'y_1 \in E$. 
Clearly, $wx_3 \notin E$ or $wx_4 \notin E$.
If $wx_3 \notin E$ and $wx_4 \notin E$ then $x_1,x_5,x_2,x_3,w,v,w',y_1$ (with center $x_1$) would induce an $S_{1,2,4}$. Thus, without loss of generality, let $wx_3 \in E$ and thus, $wx_2 \notin E$, $wx_4 \notin E$, $wx_5 \notin E$ but now, $x_1,x_2,x_5,x_4,w,v,w',y_1$ (with center $x_1$) induce an $S_{1,2,4}$ which is a contradiction.
Thus Lemma \ref{S124C3fratmostoneKwithC5} is shown. 
\qed

Recall that $K$ is a component of $G[A(v)]$, and $d \in N(v)$ contacts $K$. Let
\begin{enumerate}
\item[ ] $H := V(K) \cap N(d)$ and
\item[ ] $Z := V(K) \setminus N(d)$.
\end{enumerate}

Obviously, $\{H,Z\}$ is a partition of $V(K)$. Since $G$ is $C_3$-free, $H$ is an independent set.

For showing that MWIS can be solved for $K$ in polynomial time, let us first consider the case when $G[Z]$ is bipartite. 

\section{Case 1: $G[Z]$ is bipartite}

Recall that, if component $K$ contains no $C_5$, then by Corollary \ref{MWISS124K3C5fr}, MWIS can be solved in polynomial time for $K$.
Thus assume that $K$ contains a $C_5$, say $C$ with vertices $c_1,\ldots,c_5$ and edges $c_ic_{i+1}$ (index arithmetic modulo 5). 
Since we assume that $G[Z]$ is bipartite and since $H$ is an independent set, every $C_5$ $C$ in $K$ has at least one vertex in $H$, and thus, we have one of the following two types:

\begin{enumerate}
\item[ ] {\em Type $1$}: $C$ has exactly one vertex in $H$ (and thus, the four vertices of $C$ in $Z$ induce a $P_4$).
\item[ ] {\em Type $2$}: $C$ has exactly two vertices in $H$ (and thus, the three vertices of $C$ in $Z$ induce a $P_1+P_2$).
\end{enumerate}

\begin{fact}\label{fact0}
Let $T = (U_1, U_2, E')$ be a nontrivial component of $G[Z]$. 
If $h \in H$ contacts both sides $U_1,U_2$ of $T$ then there is a $C_5$ of type $1$ in $K$.
\end{fact}

{\bf Proof.} Let $t_1 \in U_1$ and $t_2 \in U_2$ be two neighbors of $h$. Note that $t_1$ is nonadjacent to $t_2$ since $G$ is $C_3$-free. Then, since $T$ is connected, there is a shortest path, say $P$ (of an even number of internal vertices) in $T$ between $t_1$ and $t_2$; without loss of generality, let us assume that $h$ is nonadjacent to any internal vertex of $P$ (else we may re-define the choice of $t_1$ and $t_2$). 

If $P$ has only two internal vertices, say $P=(t_1,t'_2,t'_1,t_2)$, then $(h,t_1,t'_2,t'_1,t_2)$ induce a $C_5$ of type 1 in $K$.
Thus, suppose to the contrary that $P$ has more than two internal vertices (and then $P$ has at least four internal vertices). Then $h,t_1,d,v,t_2$, and the three vertices of $P$ closest to $t_2$ induce an $S_{1,2,4}$ in $G$ which is a contradiction. Thus, Fact \ref{fact0} is shown. 
\qed 

\subsection{Case 1.1: Every $C_5$ in $K$ is of type 2.}

For $h \in H$ and nontrivial component $T = (U_1, U_2, E')$ of $G[Z]$, we define:
 
\begin{defi}\label{defihalfjoin}
\mbox{ }
\begin{itemize}
\item[$(i)$] {\em $h$ has a half-join to $T$} if either $N_T(h) = U_1$ or $N_T(h) = U_2$ $($i.e., either $h \join U_1$ and $h \cojoin U_2$ or $h \join U_2$ and $h \cojoin U_1)$.
\item[$(ii)$] {\em $h$ properly one-side contacts $T$} if either $\emptyset \subset N_T(h) \subset U_1$ or
$\emptyset \subset N_T(h) \subset U_2$.
\end{itemize}
\end{defi}

By Fact \ref{fact0} and Case 1.1, we have: 

\begin{fact}\label{fact1}
For every $h \in H$, if $h$ contacts a nontrivial component $T=(U_1,U_2,E')$ of $G[Z]$ and every $C_5$ in $K$ is of type $2$
then either $h$ has a half-join to $T$ or $h$ properly one-side contacts $T$.
\end{fact}

\begin{defi}\label{defigreencomp}
A nontrivial component $T$ of $G[Z]$ is a {\em green component of $G[Z]$} if there is a vertex $h \in H$ which properly one-side contacts $T$.
\end{defi}

{\bf Case 1.1.1} $G[Z]$ has no green component.

\begin{lemm}\label{ifnogreencomponent}
If there is no green component in $G[Z]$ then MWIS is solvable in polynomial time for $K$.
\end{lemm}

{\bf Proof.}
Since $G[Z]$ has no green component, Fact \ref{fact1} implies that, for each $h \in H$ and for each nontrivial component $T=(U_1,U_2,E')$ of $G[Z]$, if $h$ contacts $T$ then $h$ has a half-join to $T$, i.e., either $h \join U_1$ and $h \cojoin U_2$ or $h \join U_2$ and $h \cojoin U_1$. In particular, that implies: 
\begin{clai}\label{claimP4}
For each $h \in H$, there is no induced $P_3$, say $(x,y,z)$, of $G[Z]$ such that $h$ is an endpoint of the $P_4$ $(h,x,y,z)$ in $G$.
\end{clai}

For any $h \in H$ and for any $P_1 + P_2$ in $G[Z]$ with vertices $x_1,y_1,z_1$ such that $y_1z_1 \in E$ and $x_1 \cojoin \{y_1,z_1\}$, let us say that $h$ {\em doubly contacts} the $P_1 + P_2$ if $h$ is adjacent to $x_1$ and to exactly one vertex of $y_1,z_1$.

Then let $H' := \{h \in H: h$ doubly contacts a $P_1 + P_2$ in $G[Z]\}$.

If $H' = \emptyset$, then $K$ has no $C_5$ of type 2, i.e., by assumption of Case 1.1, $K$ is $C_5$-free and then, by Lemma \ref{S124C3C5frnearlybip}, MWIS can be solved in polynomial time for $K$. Thus, assume that $H' \neq \emptyset$. 

\begin{clai}\label{claimcojoin}
Let $h_1 \in H'$ and $h_2 \in H$ such that $h_1$ doubly contacts a $P_1 + P_2$ with $P_1$ $x_1$ and $P_2$ $y_1z_1$ in $G[Z]$, and $h_2$ contacts a $P_2$ $y_2z_2 \in E$ in $G[Z]$. 
If $h_2 \cojoin \{x_1,y_1,z_1\}$ then $\{y_1,z_1\} \cap \{y_2,z_2\} = \emptyset$ and $\{y_1,z_1\} \cojoin \{y_2,z_2\}$.   
\end{clai}

{\em Proof.} 
Assume without loss of generality that $h_2y_2 \in E$. Clearly, $y_2 \neq x_1,y_1,z_1$ since $h_2 \cojoin \{x_1,y_1,z_1\}$. 
By Claim \ref{claimP4} and since $G$ is $C_3$-free, $h_2,y_2,y_1,z_1$ do not induce a $P_4$ in $G$, and thus, $y_2 \cojoin \{y_1,z_1\}$ 
which implies $z_2 \neq y_1,z_1$. If $z_2=x_1$ then $z_2 \neq y_1,z_1$ and $\{y_1,z_1\} \cojoin \{y_2,z_2\}$. 
Now assume that $z_1 \neq x_1$, and recall that $z_2 \neq y_1,z_1$.

By Claim \ref{claimP4}, $h_2,y_2,z_2,y_1$ do not induce a $P_4$ in $G$, and 
correspondingly, $h_2,y_2,z_2,z_1$ do not induce a $P_4$ in $G$. Thus, $z_2 \cojoin \{y_1,z_1\}$, and Claim \ref{claimcojoin} is shown. 
$\diamond$

\medskip  
 
Now, let '$\geq$' be the following binary relation on $H'$: For any pair $h_1,h_2 \in H'$, $h_1 \geq h_2$ if either $h_1=h_2$ or $h_1$ contacts all $P_1+P_2$'s of $G[Z]$ which are doubly contacted by $h_2$. Correspondingly, $h_2 \not \geq h_1$ if vertex $h_1$ doubly contacts a $P_1 + P_2$ $P$ of $G[Z]$ such that $h_2$ does not contact $P$. 
In particular let us write $h_1 > h_2$ if $h_1 \geq h_2$ and $h_2 \not \geq h_1$.

\begin{clai}\label{claim1}
For any $h_1,h_2 \in H'$, either $h_1 \geq h_2$ or $h_2 \geq h_1$.
\end{clai}

{\em Proof.} Suppose to the contrary that $h_1 \not\geq h_2$ and $h_2 \not\geq h_1$. Then $h_1$ doubly contacts a $P_1+P_2$ of $G[Z]$ with $P_1$ $x_1$ and $P_2$ $y_1z_1$ such that $h_1$ is adjacent to $x_1,y_1$, while $h_2 \cojoin \{x_1,y_1,z_1\}$, and $h_2$ doubly contacts a $P_1+P_2$ of $G[Z]$ with $P_1$ $x_2$ and $P_2$ $y_2z_2$ such that $h_2$ is adjacent to $x_2,y_2$, while $h_1 \cojoin \{x_2,y_2,z_2\}$.

By Claim \ref{claimcojoin}, $\{y_1,z_1\} \cap \{y_2,z_2\} = \emptyset$ and $\{y_1,z_1\} \cojoin \{y_2,z_2\}$. 

Clearly, since $h_1x_1 \in E$ and $h_1 \cojoin \{x_2,y_2,z_2\}$, we have $x_1 \neq y_2$ and $x_1 \neq z_2$. 

By Claim \ref{claimP4} and since $G$ is $C_3$-free, $h_1,x_1,y_2,z_2$ do not induce a $P_4$ in $G$, which implies $x_1 \cojoin \{y_2,z_2\}$. But now, 
$h_1,x_1,y_1,z_1,d,h_2,y_2,z_2$ (with center $h_1$) induce an $S_{1,2,4}$ which is a contradiction. 
Thus, Claim \ref{claim1} is shown.
$\diamond$

\begin{clai}\label{claim2}
For any $h_1,h_2,h_3 \in H'$, if $h_1 > h_2$ and $h_2 > h_3$ then $h_1 \geq h_3$. 
\end{clai}

{\em Proof.} 
Since $h_1 > h_2$ and $h_2 > h_3$, there is a $P_1+P_2$ with $P_1$ $x_1$ and $P_2$ $y_1z_1$ in $G[Z]$ such that $h_1$ is adjacent to $x_1,y_1$, 
while $h_2 \cojoin \{x_1,y_1,z_1\}$, and there is a $P_1+P_2$ with $P_1$ $x_2$ and $P_2$ $y_2z_2$ in $G[Z]$ such that $h_2$ is adjacent to $x_2,y_2$, while 
$h_3 \cojoin \{x_2,y_2,z_2\}$.

Suppose to the contrary that $h_1 \not\geq h_3$. Then there is a $P_1+P_2$ with $P_1$ $x_3$ and $P_2$ $y_3z_3$ in $G[Z]$  
such that $h_3$ is adjacent to $x_3,y_3$ while $h_1 \cojoin \{x_3,y_3,z_3\}$. 

By Claim \ref{claimcojoin}, the sets $\{y_1,z_1\}$, $\{y_2,z_2\}$, and $\{y_3,z_3\}$ are pairwise disjoint, and $\{y_1,z_1\} \cojoin \{y_2,z_2\}$, 
$\{y_2,z_2\} \cojoin \{y_3,z_3\}$, and $\{y_1,z_1\} \cojoin \{y_3,z_3\}$.

Since $h_2 \cojoin \{x_1,y_1,z_1\}$ and $h_2x_2 \in E$, we have $z_1 \neq x_2$, and clearly, $z_1 \neq x_1$. Thus, possibly $z_1=x_3$, and analogously, 
possibly $z_2=x_1$, and $z_3=x_2$.

Now first assume that $z_1=x_3$, $z_2=x_1$, and $z_3=x_2$. Then we claim that $h_2,x_2,y_2,x_1,d,h_3,x_3,y_1$ (with center $h_2$) would induce an $S_{1,2,4}$: 

Recall that $h_2 \cojoin \{x_1,y_1,z_1\}$, $h_3 \cojoin \{x_2,y_2,z_2\}$, $\{y_1,z_1\} \cojoin \{y_2,z_2\}$, $\{y_2,z_2\} \cojoin \{y_3,z_3\}$, and 
$\{y_1,z_1\} \cojoin \{y_3,z_3\}$. Clearly, $h_2$ doubly contacts the $P_1+P_2$
with $P_1$ $x_2$ and $P_2$ $y_2x_1$. 
Then clearly, $h_3 \cojoin \{x_2,y_2,x_1\}$ and since $G$ is $C_3$-free, $x_3=z_1$ and $y_1x_3 \in E$, $h_3y_1 \notin E$.   
Moreover, $x_3 \cojoin \{h_2,x_2,x_1,y_2\}$ since $x_3=z_1$, $x_2=z_3$, $x_1=z_2$, and $\{y_1,z_1\} \cojoin \{y_2,z_2\}$.
Finally, $y_1 \cojoin \{h_2,h_3,x_1,x_2,y_2\}$ since clearly, $x_1y_1 \notin E$, $h_2 \cojoin \{x_1,y_1,z_1\}$, $y_1h_3 \notin E$ since $h_3z_1 \in E$, $z_1y_1 \in E$ and $G$ is $C_3$-free, and $x_1=z_2$, $x_2=z_3$, $\{y_1,z_1\} \cojoin \{y_2,z_2\}$, and $\{y_1,z_1\} \cojoin \{y_3,z_3\}$. 
Thus, $h_2,x_2,y_2,x_1,d,h_3,x_3,y_1$ induce an $S_{1,2,4}$ which is a contradiction, i.e., $x_3 = z_1$, $y_3 = x_1$, and $z_3 = y_1$ is impossible.

Now assume that we have exactly two such equalities. If $z_1 \neq x_3$ but $z_2 = x_1$ and $z_3 = x_2$ then we claim that 
$h_3,x_3,y_3,x_2,d,h_1,x_1,y_2$ (with center $h_3$) would induce an $S_{1,2,4}$: 

Recall that in this case, $h_3$ doubly contacts the $P_1+P_2$ with $P_1$ $x_3$ and $P_2$ $y_3x_2$, and $h_3 \cojoin \{x_2,y_2,z_2\}$, $h_1 \cojoin \{x_3,y_3,z_3\}$. Clearly, $h_1y_2 \notin E$ since $G$ is $C_3$-free.    
Since $z_2 = x_1$, $z_3 = x_2$, and $\{y_2,z_2\} \cojoin \{y_3,z_3\}$, and by Claim \ref{claimP4}, we have $x_1 \cojoin \{h_3,x_2,x_3,y_3\}$; in particular, if $x_3x_1 \in E$ then there is a $P_4$ $(h_3,x_3,x_1,y_2)$ which contradicts Claim \ref{claimP4}. Finally, $y_2 \cojoin \{h_1,h_3,x_2,y_3,x_3\}$ as before.
Thus, $h_3,x_3,y_3,x_2,d,h_1,x_1,y_2$ (with center $h_3$) induce an $S_{1,2,4}$ which is a contradiction, i.e., exactly two such equalities
$y_3 = x_1$ and $z_3 = x_2$ are impossible. 

By symmetry, we can show that the two other cases of exactly two such equalities are impossible. 

Now assume that we have exactly one such equality. By symmetry, assume that $z_1 \neq x_3$, $z_2 \neq x_1$ but $z_3 = x_2$. Then we claim that 
$h_1,x_1,y_1,z_1,d,h_2,x_2,y_3$ (with center $h_1$) would induce an $S_{1,2,4}$: 

Recall that in this case, $h_1$ doubly contacts the $P_1+P_2$ with $P_1$ $x_1$ and $P_2$ $y_1z_1$, and $h_1 \cojoin \{x_3,y_3,z_3\}$, 
$h_2 \cojoin \{x_1,y_1,z_1\}$. Clearly, $h_2y_3 \notin E$ since $G$ is $C_3$-free and $h_2x_2 \in E$, $x_2y_3 \in E$ (recall $z_3 = x_2$). 
Moreover, $\{y_1,z_1\} \cojoin \{y_3,z_3\}$. Thus, $h_1,x_1,y_1,z_1,d,h_2,x_2,y_3$ (with center $h_1$) induce an $S_{1,2,4}$ which is a contradiction, i.e., exactly one such equality is impossible. 

Finally assume that $z_1 \neq x_3$, $z_2 \neq x_1$, and $z_3 \neq x_2$. Since $h_1 > h_2$ and $h_1 \cojoin \{x_3,y_3,z_3\}$, $h_2$ does not doubly contact the $P_1+P_2$ $x_3,y_3,z_3$. 

If $h_2y_3 \in E$ (and since $G$ is $C_3$-free, $h_2z_3 \notin E$) then $h_1,x_1,y_1,z_1,d,h_2,y_3,z_3$ (with center $h_1$) would induce an $S_{1,2,4}$ 
 (recall that $\{y_1,z_1\} \cojoin \{y_3,z_3\}$, and by Claim \ref{claimP4}, we have $x_1y_3 \notin E$, $x_1z_3 \notin E$ since otherwise there is a $P_4$ $(h_1,x_1,y_3,z_3)$). 

Thus, $h_2y_3 \notin E$ and by symmetry, $h_2z_3 \notin E$. 

If $h_2x_3 \in E$ then $h_2,x_2,y_2,z_2,x_3,h_3,y_3,z_3$ (with center $h_2$) would induce an $S_{1,2,4}$ (recall $h_3 \cojoin \{x_2,y_2,z_2\}$, 
$\{y_2,z_2\} \cojoin \{y_3,z_3\}$, and by Claim \ref{claimP4}, we have $x_3y_2 \notin E$, $x_3z_2 \notin E$ since otherwise there is a $P_4$ $(h_3,x_3,y_2,z_2)$).
 
Thus, $h_2x_3 \notin E$. But then $h_3,x_3,y_3,z_3,d,h_2,y_2,z_2$ (with center $h_3$) induce an $S_{1,2,4}$ which is a contradiction. 

Thus, Claim \ref{claim2} is shown. 
$\diamond$

\begin{clai}\label{claim3}
There is a vertex $h' \in H'$ such that $h' \geq h$ for every $h \in H'$.
\end{clai}

{\em Proof.} The proof can be done by induction on the cardinality, say $k$, of $H'$. It trivially follows for $k=1$. 
If $k = 2$ then Claim \ref{claim3} follows by Claim \ref{claim1}.

Now assume that $k > 2$ and that Claim \ref{claim3} holds for $k - 1$. Let $H''$ be any subset of $k-1$ elements of $H'$, and let $Q = \{q \in H'': q \geq h$ for every $h \in H''\}$. By the inductive assumption, $Q \neq \emptyset$. Let $x \in H' \setminus H''$ (i.e., $\{x\} = H' \setminus H''$). If there is a vertex $q \in Q$ such that $q \geq x$, then $q$ is the desired vertex, and Claim \ref{claim3} follows. If there is no vertex $q \in Q$ such that $q \geq x$ then by Claim \ref{claim1}, we have $x > q$ for every $q \in Q$; on the other hand, by definition of $Q$, for every vertex $h \in H'' \setminus Q$, there is a vertex $q_h \in Q$ such that $q_h > h$. Then by Claim \ref{claim2}, we have $x \geq h$ for every $h \in H'' \setminus Q$. It implies $x \geq h$ for every $h \in H''$, i.e., $x$ is the desired vertex, and Claim \ref{claim3} is shown. 
$\diamond$

Then by repeatedly applying Claim \ref{claim3}, one can construct a total order on $H'$, say $(h_1,\ldots,h_{\ell})$, with $h_1 \geq h$ for every $h \in H' \setminus \{h_1\}$, and in general, $h_i \geq h$ for every $h \in H' \setminus \{h_1,\ldots,h_i\}$, $i \ge 2$.

Note that, by definition of $h_1$, $G[V(K) \setminus N(h_1)]$ has no $C_5$ of type 2, i.e., $G[V(K) \setminus N(h_1)]$ is $C_5$-free by assumption of Case 1. Then MWIS can be solved for $G[V(K) \setminus N(h_1)]$ in polynomial time by Lemma \ref{S124C3C5frnearlybip}.
Then MWIS can be solved on $K$ by successively solving MWIS in $G[V(K) \setminus N(h_1)]$, in
$G[(V(K) \setminus \{h_1,\ldots,h_{i-1}\}) \setminus N(h_i)]$ for $i \in \{2,\ldots,\ell\}$, and in $G[V(K) \setminus H']$. Since such graphs are $C_5$-free by construction, as shown above, this can be done in polynomial time by Corollary \ref{MWISS124K3C5fr}. This finally shows Lemma \ref{ifnogreencomponent}.
\qed

\medskip

{\bf Case 1.1.2} $G[Z]$ has a green component.

\medskip

From now on we have to assume that $G[Z]$ has green components; let $T = (U_1,U_2,E')$ be a green component of $G[Z]$. 

\begin{defi}\label{HU1outhmax}
\mbox{ }
\begin{itemize}
\item[$-$] $H_{U_1, out} := \{h \in H: h$ properly one-side contacts $T$ with respect to $U_1$, and $h$ contacts a second component $T_2$ of $G[Z]$, $T_2 \neq T\}$. 
\item[$-$] Let $h_{max} \in H_{U_1, out}$ be a vertex with maximum degree in $U_1$ over all vertices in $H_{U_1, out}$.  
\item[$-$] $Y := \{y \in U_2 \setminus N(h_{max})$: there exist $h \in H$ and $a,b \in U_1 \setminus N(h_{max})$ such that $h,a,y,b$ induce a $P_4$ in $G$ with end-vertex $h$ $($namely $h-a-y-b)\}$.
\end{itemize}
\end{defi}

{\em Remark:} $Y \neq \emptyset$ if and only if there is a vertex $h \in H$ and a component $T'$ of $G[V(T) \setminus N(h_{max})]$ such that $h$ properly one-side contacts $T'$ with respect to the $U_1$-side.

\begin{lemm}\label{lemm-maximal}
For any $y \in Y$ and for any component $T'$ of $G[(V(T) \setminus (N(h_{max}) \cup N(y))]$, no vertex of $H$
properly one-side contacts $T'$ with respect to the $U_1$-side. 
\end{lemm}

{\bf Proof.} We first show: 

\begin{clai}\label{claim1noP5}
For any $h \in H_{U_1,out}$, there are no vertices $a,b \in U_1$ and $y,w \in U_2$ such that $h,a,y,b,w$ induce a $P_5$ in $G$, namely $h-a-y-b-w$ with end-vertex $h$.
\end{clai}

{\em Proof.} Let $z$ be a neighbor of $h$ in a second component $T_2$ of $G[Z]$, and suppose to the contrary that for $a,b \in U_1$ and $y,w \in U_2$,  $h,a,y,b,w$ induce a $P_5$ in $G$, namely $h-a-y-b-w$. But then $h,z,d,v,a,y,b,w$ (with center $h$) induce an $S_{1,2,4}$ in $G$ which is a contradiction. Thus, Claim \ref{claim1noP5} is shown.
$\diamond$ 

\begin{clai}\label{claim2hout}
Let $h_1 \in H$ such that $h_1$ contacts a vertex $z \in Z \setminus V(T)$. If $h_2 \in H$ properly one-side contacts a component $T'$ of $G[V(T) \setminus N(h_1)]$ with respect to the $U_1$-side then $h_2z \in E$ and thus, $h_2 \in H_{U_1,out}$.
\end{clai}

{\em Proof.} Since $h_2$ properly one-side contacts a component $T'$ of $G[V(T) \setminus N(h_1)]$ with respect to the $U_1$-side, there exist $a,b \in U_1 \setminus N(h_1)$ and $y \in U_2 \setminus N(h_1)$ such that $h_2,a,y,b$ induce a $P_4$ $h_2-a-y-b$ in $G$. Since $h_1 \in H$ contacts a vertex $z \in Z \setminus V(T)$, i.e., $h_1z \in E$, and $d,v,h_1,z,h_2,a,y,b$ (with center $d$) do not induce an $S_{1,2,4}$ in $G$, we have $h_2z \in E$. Thus, $h_2 \in H_{U_1,out}$ and 
Claim \ref{claim2hout} is shown.  
$\diamond$ 

Assume that $Y \neq \emptyset$ since otherwise Lemma \ref{lemm-maximal} trivially follows.
Let $y \in Y$, $h \in H$ and $a,b \in U_1 \setminus N(h_{max})$ be such that $h,a,y,b$ induce a $P_4$ (namely $h-a-y-b$).
Note that by Claim \ref{claim2hout}, we have $h \in H_{U_1,out}$.
By definition of $h_{max}$, there is a vertex $x \in U_1$ such that $xh_{max} \in E$ and $xh \notin E$. 
Since $d,v,h_{max},x,h,a,y,b$ (with center $d$) do not induce an $S_{1,2,4}$, we have $xy \in E$.

Suppose to the contrary that there is a vertex $h' \in H$ such that $h'$ properly one-side contacts a component, say $T'$ of 
$G[(V(T) \setminus (N(h_{max}) \cup N(y))]$ with respect to the $U_1$-side; let $a',b' \in U_1 \setminus (N(h_{max}) \cup N(y))$ and $y' \in U_2 \setminus (N(h_{max}) \cup N(y))$ such that $h',a',y',b'$ induce a $P_4$ in $G$ (namely $h'-a'-y'-b'$).
Note that by Claim \ref{claim2hout}, $h' \in H_{U_1,out}$.

If $h' = h$ then $ha' \in E$ and $hb' \notin E$, and as above by the $S_{1,2,4}$ argument, $xy' \in E$ but then $y',b',x,y,a',h,d,v$ (with center $y'$) induce an $S_{1,2,4}$ which is a contradiction. Thus $h' \neq h$. 

Then there is a vertex $x' \in U_1$ such that $x'h_{max} \in E$ and $x'h' \notin E$. 

First assume that $x=x'$, i.e., $xh' \notin E$: Then, since $d,v,h_{max},x,h',a',y',b'$ (with center $d$) do not induce an $S_{1,2,4}$, we have $xy' \in E$. 
Since by Claim \ref{claim1noP5}, $h',a',y',x,y$ do not induce a $P_5$, we have $h'y \in E$, but then $h',y,x,y',b'$ induce a $P_5$ which is a contradiction to
 Claim \ref{claim1noP5}.

Thus, $x \neq x'$, and correspondingly, $xh' \in E$ and $x'h \in E$ (since otherwise, there is a contradiction as above for $x=x'$). 
Clearly, $h'y \notin E$ since $xh' \in E$ and $xy \in E$ and $G$ is $C_3$-free.
Since by Claim \ref{claim1noP5}, $h',a',y',x',y$ do not induce a $P_5$, we have $x'y \notin E$.

Recall that $x'x \notin E$ since $x,x' \in U_1$, and $y'y \notin E$ since $y,y' \in U_2$. 
Since $x'y \notin E$, and by Claim \ref{claim1noP5}, $h,x',y',x,y$ do not induce a $P_5$, we have $xy' \notin E$.   
 
Since $d,v,h',x,h,x',y',b'$ (with center $d$) do not induce an $S_{1,2,4}$ in $G$, we have $hb' \in E$.
But now, $d,v,h,b',h_{max},x,y,b$ (with center $d$) induce an $S_{1,2,4}$ in $G$ which is a contradiction.

Thus, Lemma \ref{lemm-maximal} is shown.
\qed \\

{\bf Case 1.1.2.1.} No vertex of $H$ properly one-side contacts two green components of $G[Z]$. \\

Let $T_1,\ldots,T_k$ denote the family of green components of $G[Z]$, and let $H_i:=\{h \in H: h$ properly one-side contacts $T_i\}$ for $i \in \{1,\ldots,k\}$.

Then, by assumption of Case 1.1.2.1 and by Fact \ref{fact1}, we have $H_i \cap H_j = \emptyset$ for $i \neq j$.\\

{\bf Case 1.1.2.1.1.} No green component of $G[Z]$ is properly one-side contacted with respect to each of its sides.\\

Then without loss of generality by symmetry, assume that every green component of $G[Z]$ is properly one-side contacted with respect to the $U_1$-side. \\

{\bf Case 1.1.2.1.1.1.} $k = 1$, i.e., there is exactly one green component. \\

Let $T = (U_1,U_2,E')$ be such a green component (i.e. $T = T_1$).

{\bf Occurrence 1.} Assume that $G[Z]$ has no other components apart from $T$. Then the vertices of $K$ are those of $H$ (which is an independent set) and of $T$. Then, by Fact \ref{fact1}, $K$ is bipartite (since the vertices of $H$ which contact $T$ can be partitioned into those contacting $U_1$ and those contacting $U_2$). Then MWIS can be solved for $K$ in polynomial time.

{\bf Occurrence 2.} Assume that $G[Z]$ has other components apart from $T$. Recall Definition~\ref{HU1outhmax} for the notions of $H_{U_1, out}$, $h_{max}$ and $Y$.  

Then one can define a total order of $H_{U_1, out}$; let us write $H_{U_1, out} = \{h_1,\ldots,h_{\ell}\}$, with $h_1 = h_{max}$, such that for $i = 2,\ldots,\ell$, vertex $h_{i}$ has maximum degree in $U_1$ over all vertices in $H_{U_1,out} \setminus \{h_1,\ldots,h_{i-1}\}$.

Note that by Lemma \ref{lemm-maximal} and by definition of $h_1 = h_{max}$, for any component $T'$ of $G[(V(T) \setminus (N(h_1) \cup N(y))]$, for any $y \in Y$, there is no vertex of $H$ which properly one-side contacts $T'$ with respect to the $U_1$-side.

Then, by assumption of Case 1.1.2.1.1, for any component $T'$ of $G[(V(T) \setminus (N(h_1)) \cup N(y))]$, for any $y \in Y$, there is no vertex of $H$ which properly one-side contacts $T'$.

Then $G[(V(K) \setminus (N(h_1) \cup N(y))]$, for any $y \in Y$, has no green component.

Furthermore, by definition of $Y$ (recall Definition~\ref{HU1outhmax}), one similarly obtains that $G[(V(K) \setminus (N(h_1) \cup Y)]$ has no green component.

Then MWIS can be solved for $G[V(K) \setminus N(h_1)]$ by successively solving MWIS for 
\begin{itemize}
\item[$(i)$] $G[(V(K) \setminus (N(h_1) \cup N(y))]$ for all $y \in Y$, and 
\item[$(ii)$] $G[(V(K) \setminus (N(h_1) \cup Y)]$. 
\end{itemize}

Since such graphs have no green component, by the above argument, this can be done in polynomial time by referring to Case 1.1.1 and Lemma \ref{ifnogreencomponent}.

Then MWIS can be solved for $K$ by successively solving MWIS for 

\begin{itemize}
\item[$(i)$] $G[V(K) \setminus N(h_1)]$,
\item[$(ii)$] $G[(V(K) \setminus (\{h_1,\ldots,h_{i-1}\} \cup N(h_i))]$ for $i \in \{2,\ldots,\ell\}$, and 
\item[$(iii)$] $G[V(K) \setminus H_{U_1, out}]$.
\end{itemize}

Concerning steps ($i$)-($ii$): such graphs have no green component, as one can check by iterating the above argument for $h_1$, so that steps ($i$)-($ii$) can be executed in polynomial time by referring to Case 1.1.1 and Lemma \ref{ifnogreencomponent}.

Concerning step ($iii$): $H \setminus H_{U_1, out}$ can be partitioned into $H' := \{h \in H \setminus H_{U_1, out}: \{h\}$ has a join either to $U_1$ or to $U_2\}$ and $H'' := H \setminus (H_{U_1, out} \cup H')$; then MWIS can be solved for $G[V(K) \setminus H_{U_1, out}]$ as follows: 

($iii.a$) solve MWIS for $G[(V(K) \setminus (H_{U_1, out} \cup N(h'))]$ for all $h' \in H'$, and 

($iii.b$) solve MWIS for $G[(V(K) \setminus (H_{U_1, out} \cup H')]$. 

By construction and by definition of $H'$, the graphs of step ($iii.a$) have no green component,  so that step ($iii.a$) can be executed in polynomial time by referring to Case 1.1.1 and Lemma~\ref{ifnogreencomponent}. 

Analogously, by construction, by Fact \ref{fact1}, and by definition of $H_{U_1, out}$ and of $H'$, the graph of step ($iii.b$) is bipartite (similarly to Occurrence 1),  so that step ($iii.b$) can be executed in polynomial time. \\

{\bf Case 1.1.2.1.1.2.} $k \geq 2$, i.e., there are at least two green components. \\

First let us prove:

\begin{fact}\label{fact2}
For every $(h_1,\ldots,h_k) \in H_1 \times \ldots \times H_k$, there is an $i \in \{1,\ldots,k\}$ such that for each $j \in \{1,\ldots,k\}$, $j \neq i$,
$h_i$ has a half-join to $T_j$.
\end{fact}

{\bf Proof.} The proof is done by induction on $k$.

Assume that $k = 2$. Then let $H' = \{h_1,h_2\}$ and let $T_1$ (respectively $T_2$) be a component of $G[Z]$ such that $h_1$ (respectively $h_2$) properly one-side contacts $T_1$ (respectively $T_2$). In particular there are: vertices $x_1,x_2,x_3 \in T_1$ inducing a $P_3$ ($x_1-x_2-x_3$) such that $h_1$ is adjacent to $x_1$, and vertices $y_1,y_2,y_3 \in T_2$ inducing a $P_3$ ($y_1-y_2-y_3$) such that $h_2$ is adjacent to $y_1$.
Suppose that the assertion is not true. Then, by Fact \ref{fact1} and since $H_1 \cap H_2 = \emptyset$, we have: $h_1$ does not contact $T_2$, and $h_2$ does not contact $T_1$. Then $d,v,h_1,x_1,h_2,y_1,y_2,y_3$ (with center $d$) induce an $S_{1,2,4}$ which is a contradiction.

Then let us assume that the assertion is true for $k-1$ and prove that it is true for $k$.
Let $(h_1,\ldots,h_k) \in H_1 \times \ldots \times H_k$. By the inductive assumption on $(h_2,\ldots,h_k)$, we can assume without loss of generality that
$h_2$ has a half-join to $T_j$ for every $j > 2$. If $h_2$ has a half-join to $T_1$ then Fact \ref{fact2} is proved.
Otherwise, by Fact \ref{fact1}, assume that $h_2 \cojoin T_1$. If for every $j > 1$, $h_1$ has a half-join to $T_j$ then Fact \ref{fact2} is proved. Otherwise, by the inductive assumption on $(h_1,h_3,\ldots,h_k)$, we can assume without loss of generality that $h_3$ has a half-join to $T_1$ and to $T_j$ for every $j > 3$. Note that $h_3$ contacts $T_2$ (and thus by Fact \ref{fact1} has an half-join to $T_2$), since otherwise $d,v,h_3$, a neighbor of $h_3$ in $T_1$ (recall that $h_3$ has a half-join to $T_1$), $h_2$, and three vertices of $T_2$ (i.e., those inducing a $P_4$ together with $h_2$) induce a $S_{1,2,4}$ (with center $d$), a contradiction. Then $h_3$ is the desired vertex, i.e., the assertion follows.

This completes the proof of Fact \ref{fact2}.  
\qed \\

Let us write $T_i=(U_{1,i},U_{2,i},E_i)$, for $i \in \{1,\ldots,k\}$.

Let us say that a vertex $h \in H_i$, for some $i \in \{1,\ldots,k\}$, is a {\em critical vertex of $K$} if
\begin{itemize}
\item[$(i)$] $h$ has maximum degree in $U_{1,i}$ over all vertices of $H_i$, and
\item[$(ii)$] for each $j \in \{1,\ldots,k\}$, $j \neq i$, $h$ has a half-join to $T_j$.
\end{itemize}

\begin{fact}\label{fact4}
There is a critical vertex of $K$.
\end{fact}

{\bf Proof.}
Let $(h_1^*,\ldots,h_k^*) \in H_1 \times \ldots \times H_k$ such that $h_i^*$ has maximum degree in $U_{1,i}$ over all vertices of $H_i$, for all
$i \in \{1,\ldots,k\}$. Then Fact \ref{fact4} follows by Fact \ref{fact2}.
\qed \\

Then let us show that, in Case 1.1.2.1.1.2, MWIS can be solved in polynomial time for $K$.

\begin{fact}\label{factMWISantineighbcritical}
For any critical vertex, say $h^*$ of $K$, MWIS can be solved in polynomial time for $G[V(K) \setminus N(h^*)]$.
\end{fact}

{\bf Proof.} By definition of a critical vertex of $K$, let $T$ be the green component of $G[Z]$, with bipartition $T = (U_1,U_2,E')$, such that $h^*$ has maximum degree in $U_1$ over all vertices of $H$ which properly one-side contact $T$ with respect to $U_1$. Then, since $h^*$ is critical and since $k \geq 2$, $h^*$ contacts a component of $G[Z]$ different to $T$ (note that in particular $h^*$ has maximum degree in $U_1$ over all vertices of $H$ which properly one-side contact $T$ with respect to $U_1$ and which contacts a component of $G[Z]$ different to $T$). Then one can apply Lemma \ref{lemm-maximal} with $h^* = h_{max}$: in particular let $Y$ be the subset of $U_2$ as in Definition~\ref{HU1outhmax} [with respect to $h^*$].

Then MWIS can be solved for $G[V(K) \setminus N(h^*)]$ by successively solving MWIS for

\begin{itemize}
\item[$(i)$] $G[(V(K) \setminus (N(h^*) \cup N(y))]$ for all $y \in Y$, and
\item[$(ii)$] $G[(V(K) \setminus (N(h^*) \cup Y)]$.
\end{itemize}

In particular, by Lemma \ref{lemm-maximal} and since $h^*$ is a critical vertex of $K$ (and by definition of $Y$), those graphs in steps $(i)-(ii)$ restricted to their intersection with $Z$ have no green component, as one can easily check by Lemma \ref{lemm-maximal}. Then steps $(i)-(ii)$ can be executed in polynomial time by referring to Case 1.1.1 which shows Fact \ref{factMWISantineighbcritical}. 
\qed 

Using Fact \ref{factMWISantineighbcritical}, MWIS can be solved in polynomial time for $K$ as follows:

Let us observe that, in view of iterating the search of critical vertices, Fact \ref{fact4} can be applied until $i \geq 2$.

Then let us write $H_{critical} = \{h_1,\ldots,h_m\} \subset H_1 \cup \ldots \cup H_k$ be such that, according to Fact \ref{fact4}, $h_i$ is a critical vertex of $G[V(K) \setminus \{h_1,\ldots,h_{i-1}\}]$ for $i \in \{1,\ldots,m\}$.

Then, as observed above, $(H_1 \cup \ldots \cup H_k) \setminus H_{critical} \subseteq H_i$ for some $i \in \{1,\ldots,k\}$.

Then MWIS can be solved for $K$ by successively solving MWIS for
\begin{itemize}
\item[$(i)$] $G[V(K) \setminus N(h_1)]$,
\item[$(ii)$] $G[(V(K) \setminus (\{h_1,\ldots,h_{i-1}\} \cup N(h_i))]$, $i \in \{2,\ldots,m\}$, and 
\item[$(iii)$] $G[V(K) \setminus H_{critical}]$.
\end{itemize}

Then, steps $(i)-(ii)$ can be executed in polynomial time by Fact \ref{factMWISantineighbcritical}, i.e., by referring to Case 1.1.1, while step $(iii)$ can be executed in polynomial time since as observed above, $G[V(K) \setminus H_{critical}]$ has exactly one green component, i.e., by referring to Case 1.1.2.1.1.1.  \\

{\bf Case 1.1.2.1.2.} A green component of $G[Z]$ is properly one-side contacted with respect to each of its sides. \\

This case can be settled similarly to Case 1.1.2.1.1. In particular (apart from Occurrence~1 which can be settled in the same way), while all subcases of Case 1.1.2.1.1 finally reduce to Case 1.1.1, all subcases of Case 1.1.2.1.2 finally reduce to Case 1.1.2.1.1. \\

{\bf Case 1.1.2.2.} There is a vertex of $H$ which properly one-side contacts at least two green components of $G[Z]$.

\begin{lemm}\label{lemmaCase1122}
For Case $1.1.2.2$, MWIS is solvable in polynomial time for component $K$.
\end{lemm}

Let $\{T_1,\ldots,T_k\}$ denote the family of green components of $G[Z]$, and let $H_i:=\{h \in H: h$ properly one-side contacts $T_i\}$ for $i \in \{1,\ldots,k\}$.
By Case 1.1.2.2, we have $k \ge 2$.

Then let $H' := H_1 \cup \ldots \cup H_k$. Clearly, $H' \neq \emptyset$ by assumption of Case 1.1.2.

Let '$\geq$' be the following binary relation on $H'$: For any pair $a,b \in H'$, $a \geq b$ if either $a=b$ or $a$ contacts all components of $G[Z]$ which are properly one-side contacted by $b$. Correspondingly, $b \not \geq a$ if $b$ does not contact all components of $G[Z]$ which are properly one-side contacted by $a$.
In particular let us write $a > b$ if $a \geq b$ and $b \not \geq a$. 

\begin{clai}\label{claim1Case1122}
For any $a,b \in H'$, either $a \geq b$ or $b \geq a$.
\end{clai}

{\em Proof.} Suppose to the contrary that $a \not \geq b$ and $b \not \geq a$. Then there is a component $T_a$ of $G[Z]$, with vertices $x_1,x_2,x_3$ inducing a $P_3$ ($x_1-x_2-x_3$), such that $a$ is adjacent to $x_1$ (and is nonadjacent to $x_2,x_3$),  while $b$ is nonadjacent to any vertex of $T_a$, and there is a component $T_b$ of $G[Z]$, with vertices $y_1,y_2,y_3$ inducing a $P_3$ ($y_1-y_2-y_3$), such that $b$ is adjacent to $y_1$ (and is nonadjacent to $y_2,y_3$), while $a$ is nonadjacent to any vertex of $T_b$. But now, $d,v,a,x_1,b,y_1,y_2,y_3$ (with center $d$) induce an $S_{1,2,4}$ which is a contradiction. Thus, Claim \ref{claim1Case1122} is shown.   
$\diamond$ 

\begin{clai}\label{claim2Case1122}
For any $a,b,c \in H'$, if $a > b$ and $b > c$ then $a \geq c$.
\end{clai}

{\em Proof.} Since $a > b$ and $b > c$, there is a component $T_a$ of $G[Z]$, with vertices $x_1,x_2,x_3$ inducing a $P_3$ $x_1-x_2-x_3$ such that $ax_1 \in E$ (and $ax_2 \notin E, ax_3 \notin E$),  while $b$ is nonadjacent to any vertex of $T_a$, and there is a component $T_b$  of $G[Z]$ with vertices $y_1,y_2,y_3$ inducing a $P_3$ $y_1-y_2-y_3$ such that $by_1 \in E$ (and $by_2 \notin E,by_3 \notin E$), while $c$ is nonadjacent to any vertex of $T_b$.

Suppose to the contrary that $a \not \geq c$. Then there is a component $T_c$ of $G[Z]$, with vertices $z_1,z_2,z_3$ inducing a $P_3$ $z_1-z_2-z_3$ such that 
$cz_1 \in E$ (and $cz_2 \notin E,cz_3 \notin E$), while $a$ is nonadjacent to any vertex of $T_c$.

Since $d,v,c,z_1,b,y_1,y_2,y_3$ (with center $d$) do not induce an $S_{1,2,4}$, we have $bz_1 \in E$, and
since $b,y_1,z_1,z_2,d,a,x_1,x_2$ (with center $b$) do not induce an $S_{1,2,4}$, we have $ay_1 \in E$.
But now, $d,v,a,y_1,c,z_1,z_2,z_3$ (with center $d$) induce an $S_{1,2,4}$ which is a contradiction.

Thus, Claim \ref{claim2Case1122} is shown. 
$\diamond$

\begin{clai}\label{claim3Case1122}
There is a vertex $h' \in H'$ such that $h' \geq h$ for every $h \in H'$.
\end{clai}

{\em Proof.} The proof is done by induction on the cardinality, say $k$, of $H'$. It trivially follows for $k=1$.
If $k = 2$ then Claim \ref{claim3Case1122} follows by Claim \ref{claim1Case1122}.

Now assume that $k > 2$ and that Claim \ref{claim3Case1122} holds for $k - 1$. Let $H''$ be any subset of $k-1$ elements of $H'$. Let $Q := \{q \in H'': q \geq h$ for every $h \in H''\}$. By the inductive assumption we have $Q \neq \emptyset$. Let $x \in H' \setminus H''$ (i.e., $\{x\} = H' \setminus H''$). If there is a vertex $q \in Q$ such that $q \geq x$, then $q$ is the desired vertex, and the claim follows. If there is no vertex $q \in Q$ such that $q \geq x$, then by Claim \ref{claim1Case1122}, we have $x > q$ for every $q \in Q$; on the other hand, by definition of $Q$, for every vertex $h \in H'' \setminus Q$ there is a vertex $q_h \in Q$ such that $q_h > h$; then, 
by Claim \ref{claim2Case1122}, we have $x \geq h$ for every $h \in H'' \setminus Q$. Thus, $x \geq h$ for every $h \in H''$, that is $x$ is the desired vertex, and Claim \ref{claim3Case1122} is shown. 
$\diamond$

Let us say that a vertex $h' \in H'$ is {\em basic for $H'$} if
\begin{itemize}
\item[$(i)$] $h' \geq h$ for every $h \in H'$, and
\item[$(ii)$] $h'$ has maximum degree in $Z$ over all vertices enjoying $(i)$ 
\end{itemize} 
 
Thus, if there is a vertex $h'' \in H'$ which enjoys $(i)$ and if $h''$ has a neighbor $z'' \in Z$ being nonadjacent to $h'$, then $h'$ has a neighbor $z' \in Z$ being nonadjacent to $h''$.

Note that by Claim \ref{claim3Case1122}, there is a basic vertex for $H'$.  

\begin{clai}\label{claim4Case1122}
Let $h' \in H'$ be a basic vertex for $H'$. Then no vertex of $H$ properly one-side contacts two components of $G[Z \setminus N(h')]$.
\end{clai}

{\em Proof.} Suppose to the contrary that there is a vertex $h \in H$ such that $h$ properly one-side contacts two components of $G[Z \setminus N(h')]$. Then $h \in H'$ (since $h$ properly one-side contacts at least one component of $G[Z]$). Then let $x_1,x_2,x_3$ and $y_1,y_2,y_3$ be respectively vertices of such components, inducing $P_3$'s $x_1-x_2-x_3$ and $y_1-y_2-y_3$, with $hx_1 \in E, hy_1 \in E$. Since $h'$ is basic, $h'$ has a neighbor $z' \in Z$ such that $hz' \notin E$ (either by $(i)$ or by $(ii)$ of the definition of a basic vertex).

If $z'$ does not contact either $\{x_1,x_2,x_3\}$ or $\{y_1,y_2,y_3\}$, say $\{z'\} \cojoin \{x_1,x_2,x_3\}$ without loss of generality by symmetry, then $d,v,h',z',h,x_1,x_2,x_3$ (with center $d$) induce an $S_{1,2,4}$ which is a contradiction.

Thus assume that $z'$ contacts $\{x_1,x_2,x_3\}$ as well as $\{y_1,y_2,y_3\}$. Then $z',x_1,x_2,x_3,y_1,y_2,y_3$ belong to the same component of $G[Z]$, say $T$. 
By Case 1.1.2.2, there exists another component of $G[Z]$, say $T_+$, which is properly one-side contacted by some vertex of $H$. 

By definition of $h'$, vertex $h'$ contacts $T_+$; let $t \in T_+$ be adjacent to $h'$. Then assume that $t$ is adjacent to $h$ as well (since, otherwise, one can apply the previous $S_{1,2,4}$ argument with $t$ instead of $z'$). Then, by symmetry, let us consider the following exhaustive cases.

If $z'x_1 \in E$ then, since $G$ is $C_3$-free, $z'x_2 \notin E$.
Then $z'x_3 \in E$ since $h',t,d,v,z',x_1,x_2,x_3$ (with center $h'$) do not induce an $S_{1,2,4}$. 
Similarly, if $z'x_3 \in E$ then it follows that $z'x_1 \in E$.

Furthermore, since $z'$ contacts $\{y_1,y_2,y_3\}$, by the above and by a similar argument to the previous one, if $z'y_1 \in E$ or $z'y_3 \in E$ then 
$z'y_1 \in E$ and $z'y_3 \in E$. But now, $h,t,d,v,y_1,z',x_1,x_2$ (with center $h$) induce an $S_{1,2,4}$ which is a contradiction.

Assume that $z'$ is adjacent to $x_2,y_1,y_3$ (and then clearly, $z'$ is nonadjacent to $x_1,x_3,y_2$). But now, $h,t,d,v,y_1,z',x_2,x_3$ (with center $h$) induce an $S_{1,2,4}$ which is a contradiction.

Finally, assume that $z'$ is adjacent to $x_2,y_2$ (and then clearly, $z'$ is nonadjacent to $x_1,x_3,y_1,y_3$). But now, $h,t,d,v,x_1,x_2,z',y_2$ (with center $h$) induce an $S_{1,2,4}$ which is a contradiction.

Thus, Claim \ref{claim4Case1122} is shown. 
$\diamond$

Then by repeatedly applying Claim \ref{claim3Case1122}, one can define a total order on $H'$, say $H' = (h_1,\ldots,h_{\ell})$, such that $h_1$ is basic for $H'$, $h_2$ is basic for $H' \setminus \{h_1\}$, and so on.

Note that, by definition of $h_1$ and by Claim \ref{claim4Case1122}, there is no vertex of $H$ which properly one-side contacts two (green) components of $G[Z \setminus N(h_1)]$. Then MWIS can be solved for $G[V(K) \setminus N(h_1)]$ in polynomial time referring to Case 1.1.2.1.
Then MWIS can be solved on $K$ by successively solving MWIS in $G[V(K) \setminus N(h_1)]$, in
$G[(V(K) \setminus \{h_1,\ldots,h_{i-1}\}) \setminus N(h_i)]$ for $i \in \{2,\ldots,\ell\}$, and in $G[V(K) \setminus H']$. Since such graphs enjoy Case 1.1.2.1, this can be done in polynomial time by referring to Case 1.1.2.1.
This finally shows Lemma \ref{lemmaCase1122}.
\qed

\subsection{Case 1.2: $K$ contains a $C_5$ of type 1.}

For any $C_5$ of type 1 in component $K$, say $C$ with vertex set $V(C)=\{v_1, \ldots, v_5\}$ and edges $v_iv_{i+1}$ (index arithmetic modulo 5) such that
$V(C) \cap H =\{v_5\}$, let us say that $v_5$ is the {\em nail} $h=v_5$ of $C$, and the other vertices of $C$ are the {\em non-nail vertices} of $C$.
Then let
\begin{itemize}
\item[ ] $L(h) := \{z \in Z: z$ belongs to a $C_5$ of type 1 in $K$ with nail $h$, and $zh \notin E\}$.
\end{itemize}
Note that $v_2,v_3 \in L(h)$.

\begin{fact}\label{fact6}
Let $h^* \in H$ be such that $h^*$ has maximum degree in $Z$ over the vertices of $H$. Let $C$ be a $C_5$ of type $1$ in $G[K \setminus N(h^*)]$ with vertex set $V(C)=\{v_1, \ldots, v_5\}$ and edges $v_iv_{i+1}$ $($index arithmetic modulo $5)$ and with nail $h=v_5$. Then for every $C_5$ $C'$ of type~$1$ in $G[V(K) \setminus (N(h^*) \cup N(h))]$ we have:
\begin{itemize}
\item[$(i)$] $N(v_2) \cap V(C') \neq \emptyset$ and $N(v_3) \cap V(C') \neq \emptyset$.
\item[$(ii)$] $L(h) \cap V(C') \neq \emptyset$.
\end{itemize}
\end{fact}

{\bf Proof.}
Let $C$ be a $C_5$ of type 1 in $G[K \setminus N(h^*)]$ as described in Fact \ref{fact6} and let $C'$ be a $C_5$ of type 1 in $G[V(K) \setminus (N(h^*) \cup N(h))]$, say, with vertex set $V(C')=\{u_1, \ldots, u_5\}$ and edges $u_iu_{i+1}$ (index arithmetic modulo 5), such that $V(C') \cap H =\{u_5\}$, and let us show that statements $(i)$ and $(ii)$ hold.
Clearly $h=v_5 \notin V(C')$ and $v_1,v_4 \notin V(C')$ since $V(C') \cap N(h)=\emptyset$.

First assume that $\{u_1, \ldots, u_5\} \cap \{v_1, \ldots, v_5\} \neq \emptyset$. Then by the above, we have $\{u_1, \ldots, u_5\} \cap \{v_2,v_3\} \neq \emptyset$ which clearly means that $(i)$ and $(ii)$ hold.

Thus, from now on, assume that $\{u_1, \ldots, u_5\} \cap \{v_1, \ldots, v_5\}=\emptyset$.

Since by assumption of Case 1, $G[Z]$ is bipartite, we can assume without loss of generality that $v_1,v_3,u_1,u_3$ form an independent set, say $v_1,v_3,u_1,u_3$ are $black$, and similarly, $v_2,v_4,u_2,u_4$ form an independent set, say $v_2,v_4,u_2,u_4$ are $white$.

Since $h^*$ has maximum degree in $Z$ and since $h = v_5$ has a neighbor in $Z$, namely $v_1$, which is nonadjacent to $h^*$, there exists a neighbor of $h^*$ in $Z$, say $z$, which is nonadjacent to $h = v_5$. In particular let us assume without loss of generality that $z$ is white. We first claim: 
\begin{equation}\label{zadjv1v3}
zv_1 \in E \mbox{ and } zv_3 \in E.
\end{equation}

{\em Proof.}
Recall that $zv_5 \notin E$. Since $d,v,h^*,z,v_5,v_4,v_3,v_2$ (with center $d$) do not induce an $S_{1,2,4}$, we have $zv_3 \in E$, and
since $v_3,v_4,v_2,v_1,z,h^*,d,v$ (with center $v_3$) do not induce an $S_{1,2,4}$, we have $zv_1 \in E$. 
$\diamond$

Next we claim:
\begin{equation}\label{znonadju5}
zu_5 \notin E.
\end{equation}

{\em Proof.}
Suppose to the contrary that $zu_5 \in E$. Then, since $G$ is $C_3$-free, $zu_1 \notin E$ and $zu_4 \notin E$, and since $z$ and $u_2$ are white, $zu_2 \notin E$.

Since $u_3,u_4,u_2,u_1,z,h^*,d,v$ (with center $u_3$) do not induce an $S_{1,2,4}$, we have $zu_3 \notin E$.

Then, since $z,h^*,v_1,v_5,u_5,u_1,u_2,u_3$ (with center $z$) do not induce an $S_{1,2,4}$, we have $v_1u_2 \in E$.

Recall $v_5u_1 \notin E$, $v_5u_2 \notin E$, $v_5u_3 \notin E$, and since $v_1,u_1,u_3$ are black, we have $v_1u_1 \notin E$, and $v_1u_3 \notin E$.
Then, since $u_2,u_1,u_3,u_4,v_1,z,h^*,d$ (with center $u_2$) do not induce an $S_{1,2,4}$, we have $v_1u_4 \in E$.

But then  $v_1,u_4,u_2,u_1,z,h^*,d,v$ (with center $v_1$) induce an $S_{1,2,4}$ which is a contradiction. Thus, (\ref{znonadju5}) is shown.
$\diamond$

Then by (\ref{zadjv1v3}) and symmetry, $zu_1 \in E$ and $zu_3 \in E$. 

Moreover, we claim:
\begin{equation}\label{u5nonadjv1}
u_5v_1 \notin E.
\end{equation}

{\em Proof.} If $u_5v_1 \in E$ then, since $G$ is $C_3$-free, $v_1u_4 \notin E$ but then $z,u_1,u_3,u_4,v_1,v_5,d,v$ (with center $z$) induce an $S_{1,2,4}$. Thus, $u_5v_1 \notin E$.
$\diamond$
 
Next we claim:
\begin{equation}\label{v4adju1u3}
v_4u_1 \in E \mbox{ and } v_4u_3 \in E.
\end{equation}

{\em Proof.} Since $d,v,v_5,v_4,h^*,z,u_1,u_2$ (with center $d$) do not induce an $S_{1,2,4}$, we have $v_4u_1 \in E$. 
Now, since $d,v,v_5,v_4,h^*,z,u_3,u_2$ (with center $d$) do not induce an $S_{1,2,4}$, we have $v_4u_3 \in E$. 
$\diamond$
 
Then $u_3 \in L(h) \cap V(C')$ since $h,v_1,z,u_3,v_4$ induce a $C_5$ with nail $h$. This implies statement $(ii)$ of Fact \ref{fact6}.

Recall that by (\ref{u5nonadjv1}), $u_5v_1 \notin E$. Then we claim:
\begin{equation}\label{v1adju2}
v_1u_2 \in E.
\end{equation}

{\em Proof.} Since $d,v,v_5,v_1,u_5,u_1,u_2,u_3$ do not induce an $S_{1,2,4}$, and since $u_5v_1 \notin E$, we have $v_1u_2 \in E$. 
$\diamond$

Finally we claim:
\begin{equation}\label{v3adju4oru5andv2adju1oru3}
(v_3u_4 \in E \mbox{ or } v_3u_5 \in E) \mbox{ and } (v_2u_1 \in E \mbox{ or } v_2u_3 \in E \mbox{ or } v_2u_5 \in E).
\end{equation}

{\em Proof.} Since $d,v,u_5,u_4,h^*,z,v_3,v_4$ (with center $d$) do not induce an $S_{1,2,4}$, we have $v_3u_4 \in E$ or $v_3u_5 \in E$. 
Since $u_2,u_3,v_1,v_2,u_1,u_5,d,v$ (with center $u_2$) do not induce an $S_{1,2,4}$, we have $v_2u_1 \in E$ or $v_2u_3 \in E$ or $v_2u_5 \in E$. 
$\diamond$

Proposition (\ref{v3adju4oru5andv2adju1oru3}) implies statement $(i)$ of Fact \ref{fact6}.

Thus, Fact \ref{fact6} is shown. 
\qed 

\begin{fact}\label{fact7}
Let $h \in H$ be the nail of a $C_5$ of type $1$ in $G[V(K) \setminus N(h^*)]$. Then MWIS can be solved in polynomial time for $G[V(K) \setminus (N(h^*) \cup N(h))]$.
\end{fact}

{\bf Proof.} MWIS can be solved for $G[V(K) \setminus (N(h^*) \cup N(h)]$ by solving MWIS for
\begin{enumerate}
\item[$(i)$] $G[V(K) \setminus (N(h^*) \cup N(h) \cup N(x))]$ for any $x \in L(h)$, and
\item[$(ii)$] $G[V(K) \setminus (N(h^*) \cup N(h) \cup L(h))]$.
\end{enumerate}

Note that by Fact \ref{fact6} $(i)$, the subgraphs $G[V(K) \setminus (N(h^*) \cup N(h) \cup N(x))]$ for $x \in L(h)$ contain no $C_5$ of type 1, and by Fact \ref{fact6} $(ii)$ and by definition of $L(h)$, subgraph $G[V(K) \setminus (N(h^*) \cup N(h) \cup L(h))]$ contains no $C_5$ of type 1. Then steps $(i)-(ii)$ can be executed in polynomial time by referring to Case 1.1.
\qed \\

Then in Case 1.2, MWIS can be solved for $K$ in polynomial time as follows:

Let us write $H = \{h_1,h_2,\ldots,h_p\}$ and let us assume without loss of generality that the degree of $h_i$ in $Z$ is greater than or equal to the degree of $h_{i+1}$ in $Z$ (for $i = 1,\ldots,p-1$).

Then let $A_i := \{a \in H: a$ is the nail of a $C_5$ of type 1 in $G[V(K) \setminus (N(h_1) \cup \ldots \cup N(h_i))]\}$ (for $i = 1,\ldots,p-1$).

Then MWIS can be solved for $K$ by successively solving MWIS in $G[V(K) \setminus N(h_i)]$, for $i = 1, \ldots, p$.

In particular, for any fixed $i$, this can be done by solving MWIS in polynomial time for
\begin{enumerate}
\item[$(i)$] $G[V(K) \setminus (N(h_i) \cup N(a))]$, for every $a \in A_i$, by Fact \ref{fact7};
\item[$(ii)$] $G[V(K) \setminus (N(h_i) \cup A_i)]$, which contains no $C_5$ of type 1, by referring to Case 1.1.
\end{enumerate}

\section{Case 2: $G[Z]$ is not bipartite}

By Lemma \ref{S124C3C5frnearlybip}, we can focus on $C_5$ for odd cycles in $G[Z]$. By Lemma \ref{S124C3fratmostoneKwithC5}, we have: 

\begin{fact}\label{fact8}
If $G[Z]$ contains a $C_5$ then there is exactly one component of $G[Z]$ which contains a $C_5$.
\end{fact}

According to Lemma \ref{S124C3fratmostoneKwithC5}, let $Z^*$ be the unique component of $G[Z]$ which is not bipartite, and let $H^* := \{h \in H:h$ contacts $Z^*\}$.

\begin{fact}\label{fact9}
For every $h \in H^*$, $G[Z^* \setminus N(h)]$ is bipartite.
\end{fact}

{\bf Proof.} Since $G$ is ($S_{1,2,4},C_3$)-free, $h$ contacts every odd chordless cycle in $G[Z^*]$ (else there would be an $S_{1,2,4}$ in a subgraph with $v,d,h$, a shortest path between $h$ and a vertex, say $x_1$ in a $C_{2k+1}$ $(x_1,\ldots,x_{2k+1})$, $k \ge 2$, as well as, without loss of generality, $x_{2k+1},x_2,x_3$).
\qed \\

Then in Case 2, MWIS can be solved for $K$ in polynomial time as follows:

According to the notation above, MWIS can be solved for $K$ by successively solving MWIS for
\begin{enumerate}
\item[$(i)$] $G[V(K) \setminus N(h)]$ for every $h \in H^*$;
\item[$(ii)$] $G[V(K) \setminus H^*]$.
\end{enumerate}

Concerning step $(i)$: It can be executed in polynomial time by Facts \ref{fact8} and \ref{fact9}, i.e., by referring to Case 1 (i.e., when $G[Z]$ is bipartite).

Concerning step $(ii)$: According to Fact \ref{fact8}, $G[V(K) \setminus H^*]$ is partitioned into components, namely $Z^*$ and (possibly) other components $Q$ such that $G[V(Q) \cap Z]$ is bipartite.

Concerning $G[Z^*]$, MWIS can be solved in polynomial time for $G[Z^*]$ as follows:
\begin{enumerate}
\item[$(a)$] fix any vertex $h^* \in H^*$
\item[$(b)$] solve MWIS for $G[Z^*]$ by referring to Case 1.
\end{enumerate}
In fact, $Z^*$ can be partitioned into independent set $Z^* \cap N(h^*)$ and $Z^* \setminus N(h^*)$ (by Fact \ref{fact9}, $G[Z^* \setminus N(h^*)]$ is bipartite). Concerning the other components $Q$, MWIS can be solved in polynomial time for $G[Q]$ by referring to Case 1.
\qed

Summarizing the previous results, we have:

\begin{theo}
The MWIS problem can be solved in polynomial time for $(S_{1,2,4},C_3)$-free graphs.
\end{theo}

\section{Conclusion}

In this paper, we have shown that MWIS can be solved for $(S_{1,2,4},C_3)$-free graphs in polynomial time (the time bound of our solution algorithm may be estimated as ${\cal O}(n^{16})$). By the solution method described in Section 3, it is not difficult to derive the following result:

\begin{theo}
For every $(S_{1,2,4},C_3)$-free graph $G$ there is a family ${\cal S}$ of subsets of $V(G)$ inducing bipartite subgraphs of $G$, with ${\cal S}$ detectable in polynomial time and containing polynomially many members, such that every maximal independent set of $G$ is contained in some member of ${\cal S}$.
\end{theo}

Recall that a graph is {\em prime} if it admits no proper (non-trivial) vertex subset $U$ such that all vertices of $U$ are adjacent to the same vertices outside of $U$. The main result of this paper can be extended in various ways as follows:

{\bf Remark}. Let us recall two results by Olariu:
\begin{enumerate}
\item[$(i)$] Every paw-free graph is either $C_3$-free or complete multipartite \cite{Olari1988}.
\item[$(ii)$] If a prime graph contains a $C_3$ then it contains a house, bull, or double-gem \cite{Olari1990}.
\end{enumerate}

It is well known that MWIS can be reduced to prime graphs (see e.g.\ \cite{LozMil2008}). The main result of our paper implies that MWIS can be solved for ($S_{1,2,4}$, paw)-free graphs in polynomial time directly by $(i)$, and that more generally MWIS can be solved for ($S_{1,2,4}$, house, bull, double-gem)-free graphs in polynomial time by $(ii)$ and by results from modular decomposition theory (see e.g.\ \cite{BraLeSpi1999,McCSpi1999,MoeRad1984/1}). 

Recall that recently, Maffray and Pastor \cite{MafPas2016} showed that MWIS can be solved in polynomial time for ($S_{1,2,3}$,bull)-free graphs (after a corresponding previous result of them about MWIS for ($P_6$,bull)-free graphs \cite{MafPas2015}).

{\bf Question:} What is the complexity of MWIS for ($S_{1,2,4}$,bull)-free graphs?

\medskip

\noindent
{\bf Acknowledgment.} The second author would like to witness that he just tries to pray a lot and is not able to do anything without that - ad laudem Domini.
 
\begin{footnotesize}
\renewcommand{\baselinestretch}{0.4}

\end{footnotesize}

\end{document}